\documentclass{emulateapj}
\usepackage{graphicx}

\slugcomment{Accepted for publication in the Astrophysical Journal, March 23, 2010.}
\shorttitle{The Formation of the Haumea System}
\shortauthors{Leinhardt, Marcus, and Stewart}

\usepackage{natbib}
\makeatletter

\def\compoundrel#1\over#2{\mathpalette\compoundreL{{#1}\over{#2}}}
\def\compoundreL#1#2{\compoundREL#1#2}
\def\compoundREL#1#2\over#3{\mathrel
      {\vcenter{\hbox{$\m@th\buildrel{#1#2}\over{#1#3}$}}}}
\makeatother
\begin{document} 

\title{The Formation of the Collisional Family around the Dwarf Planet
  Haumea}

\author{Zo\"e M. Leinhardt}
\affil{Department of Applied Mathematics and Theoretical Physics,\\ 
	University of Cambridge, CB3 0WA, U.K.}
\email{Z.M.Leinhardt@damtp.cam.ac.uk}

\author{Robert A. Marcus}
\affil{Astronomy Department, Harvard University, \\
	60 Garden St., Cambridge, MA 02138, U.S.A.}

\author{Sarah T. Stewart}
\affil{Department of Earth and Planetary Sciences, Harvard University, \\
  20 Oxford St., Cambridge, MA 02138, U.S.A.}

\begin{abstract}
  Haumea, a rapidly rotating elongated dwarf planet ($\sim 1500$ km in diameter), has two
  satellites and is associated with a ``family" of several smaller
  Kuiper Belt objects (KBOs) in similar orbits.  All members of the
  Haumea system share a water ice spectral feature that is distinct
  from all other KBOs.  The relative velocities between the Haumea
  family members are too small to have formed by catastrophic
  disruption of a large precursor body, which is the process that
  formed families around much smaller asteroids in the Main Belt.
  Here we show that all of the unusual characteristics of the Haumea
  system are explained by a novel type of giant collision: a
  graze-and-merge impact between two comparably sized bodies. The
  grazing encounter imparted the high angular momentum that spun off
  fragments from the icy crust of the elongated merged body. The
  fragments became satellites and family members.  Giant collision
  outcomes are extremely sensitive to the impact parameters. Compared
  to the Main Belt, the largest bodies in the Kuiper Belt are more
  massive and experience slower velocity collisions; hence, outcomes
  of giant collisions are dramatically different between the inner and
  outer solar system. The dwarf planets in the Kuiper Belt record an
  unexpectedly large number of giant collisions, requiring a special dynamical
  event at the end of solar system formation.

\end{abstract}


\section{Introduction}
The four largest dwarf planets in the Kuiper Belt form a distinct
population of bodies with high albedos and volatile-rich surfaces
\citep{Schaller2007,Stansberry2008}.  A significant history of
collisions is suggested by the abundance of satellites in this group,
which is much higher than expected for the Kuiper Belt as a whole
\citep{Brown2006}.  Three of the four have known satellites: Pluto has
three, Haumea (formerly 2003 EL$_{61}$) has two, Eris (2003
UB$_{313}$) has one, and Makemake (2005 FY$_9$) has no substantial
satellite \citep{Brown2008}.  The size and orbits of these satellites
are different from those found around smaller (100-km size) Kuiper
Belt Objects (KBOs). To date, most known satellites around smaller KBOs are thought to have formed via a still-debated capture mechanism \citep{Knoll2008}. Hence, a different
satellite formation process is needed for the dwarf planets, and the
most promising mechanism is collisions. Recently, numerical
simulations support a giant collision origin for Pluto's massive
satellite, Charon \citep{Canup2005}. However, the formation of the
smaller satellites on the other dwarf planets has not been studied in
detail.

Haumea, a $\sim 1500$ km diameter classical belt object with a semi-major axis of 43 AU, is a particularly puzzling case as it is also associated with
several smaller KBOs with diameters between 70 and 365 km.  The smaller KBOs share similar orbits and surface
properties. The associated KBOs have been likened to collisionally-produced dynamically and compositionally associated ``families" that are observed in the asteroid belt \citep{Brown2007}. We collectively refer to Haumea, its satellites and proposed family members as the Haumea system.
Haumea has the only known family in the Kuiper Belt.  The Haumea family members share a deep
water spectral feature and neutral color
\citep{Brown2007,Ragozzine2007,Schaller2008}. The water feature is
unique in the Kuiper Belt \citep{Brown2007} and indicative of
unusually carbon-free water ice \citep{Pinilla2009}.

Haumea also has the distinction of being the only known highly elongated dwarf planet though its precise shape is not known. It has a spin period of only 3.9 hours \citep{Rabinowitz2006},
the fastest of all the major and dwarf planets.  The surface of Haumea
is nearly homogenous with the exception of a red spot or faint red
hemisphere \citep{Lacerda2008,Lacerda2009}; hence, the light curve is primarily a
reflection of the non-spherical shape. Using the observed light curve
and rotation period, \citet{Rabinowitz2006} fit a density of 2.6
g~cm$^{-3}$ assuming an equilibrium fluid body (a Jacobi ellipsoid).
Although the dimensions of Haumea are not yet uniquely constrained
\citep{Lockwood2009}, the observations require a tri-axial
shape (see Table~\ref{tab:sims}) \citep{Rabinowitz2006}. The derived
density is greater than the average of $\sim 2$ g~cm$^{-3}$ for the
largest KBOs \citep{Brown2008}, although the density may be smaller
with some internal friction \citep{Holsapple2007}.

Based on the relatively clean water ice surface and higher than
average bulk density, \citet{Brown2007} argue that Haumea is
differentiated with a rocky core and icy mantle. They propose that the
family members and satellites are collisionally-derived fragments that
originated primarily from the icy mantle.  The satellites and family
members are orders of magnitude less massive than Haumea, and the
family members have a minimum velocity at infinity ($V_{\infty}$) of
about 150 m~s$^{-1}$ \citep{Ragozzine2007,Ragozzine2009}. The velocity
dispersion is much less than expected if the Haumea family formed as
the result of a catastrophic impact, as in the formation of asteroid
belt families \citep[e.g., ][]{Nesvorny2006,Michel2004}. In a catastrophic disruption
event, a parent body is disrupted and dispersed such that the largest
remnant is less than or equal to half the original mass. In the gravity regime, the
fragments have initial velocities relative to the largest remnant
comparable to the escape speed ($V_{\rm esc}$) of the disrupted parent
body \citep{Benz1999}. Thus, the observed magnitude of the velocity
dispersion ($V_{\infty}$) of asteroid belt families is significant
with respect to $V_{\rm esc}$ of the largest remnant; in other words,
$V_{\rm esc}$ of the parent body was much greater than $V_{\rm esc}$
of the largest remnant in gravity dominated disruption events. Unlike most asteroid belt families, the velocity
dispersion among Haumea family members is a small fraction of the
escape velocity from Haumea ($V_{\rm esc} \sim
900$~m~s$^{-1}$). Based on the current models of family formation via catastrophic disruption, the Haumea family could not have formed
by catastrophic disruption of a much larger parent body.  

Two impact scenarios have been proposed for the formation of the
Haumea system.  \citet{Brown2007} proposed an impact event that falls
in the catstrophic disruption category, which does not agree with the
observed velocity dispersion among family members. To explain the
small velocity dispersion, \cite{Schlichting2009} suggest the breakup
of a single large moon in orbit around Haumea. However, they do not
provide an explanation for the initial state: a large moon in close
orbit around a fast-spinning, elongated planet.  To date, no known
impact scenario explains all of the unusual characteristics of the
Haumea system.

In this work, we quantitatively model the formation of the Haumea
system. We propose that the Haumea family formed via a novel type of
giant collision: a {\it graze and merge} impact between two comparably
sized bodies resulting in high angular momentum, which spun off icy
fragments that became satellites and family members. The analytic and
numerical methods are described in \S \ref{sec:method}. The results
are presented in \S \ref{sec:results}, and the implications for giant
impacts in the Kuiper Belt are discussed in \S \ref{sec:disc}.

\section{Methods} \label{sec:method}

To reduce the parameter space of possible collisions that produce a
Haumea-like system, we used a three step process: 1) derive an
analytic prediction of plausible impact parameters (\S \ref{sec:ana});
2) conduct low-resolution simulations over a broad parameter space
based on the results of the analytic prediction (\S \ref{sec:lowres});
and 3) simulate the most promising impact scenarios in high
resolution (\S \ref{sec:gadget}). The next three sections outline the
method used in each step.

\subsection{Analytic Prediction of Impact Parameters}\label{sec:ana}

Following \cite{Canup2001}, using the conservation of energy and
momentum, we derived an expression for the impact parameter and
projectile-to-target mass ratio needed to obtain the observed angular
momentum of Haumea via a giant impact.  Because Haumea is rotating
near its spin instability limit and the observed velocity dispersion
among family members is small, we consider the case where all of the
angular momentum from the collision is retained in the remaining body.  Assuming that the relative velocity between the
projectile and target was zero at infinity, the impact parameter, $b$,
is given by
\begin{equation}
b = \frac{L}{L_{crit}} \frac{k}{\sqrt{2} f(\gamma)}, \label{eq:final}
\end{equation}
where $b$ is in units of the sum of the radii of the projectile and target,
$L$ is the angular momentum, and $L_{crit} = k M^{5/3} G^{1/2} (\frac{3}{4 \pi \rho})^{1/6}$, where $k$ is the inertial constant ($2/5$ for a sphere), $M$ is the total mass, and $\rho$ is the bulk density. $L_{crit}$ is, therefore, the critical spin angular momentum that a spherical body with constant density can sustain. The mass ratio of projectile to the total mass, $\gamma = \frac{M_P}{M_T+M_P}$, enters Eq.~\ref{eq:final} through $f(\gamma) = \gamma(1-\gamma)(\gamma^{1/3} + (1 - \gamma^{1/3}))^{1/2} \sin \theta$ where $\theta$ is the impact angle.  However, if we assume that the impact velocity is
greater than $V_{\rm esc}$, the total energy equation will
no longer equal zero \citep[Eq.~B3 of][]{Canup2001}. In this case,
the impact parameter is constrained by
\begin{equation} 
  b = \frac{L}{L_{crit}} \frac{k}{\sqrt{2} f(\gamma)} \frac{V_{\rm{esc}}}{V_i}, 
  \label{eq:bvi}
\end{equation}
where $V_i$ is the impact velocity and $V_{\rm esc}$ is the mutual
escape velocity \citep[Eq.~1 of][]{Canup2005}.

\begin{figure}
 \includegraphics[scale=0.43]{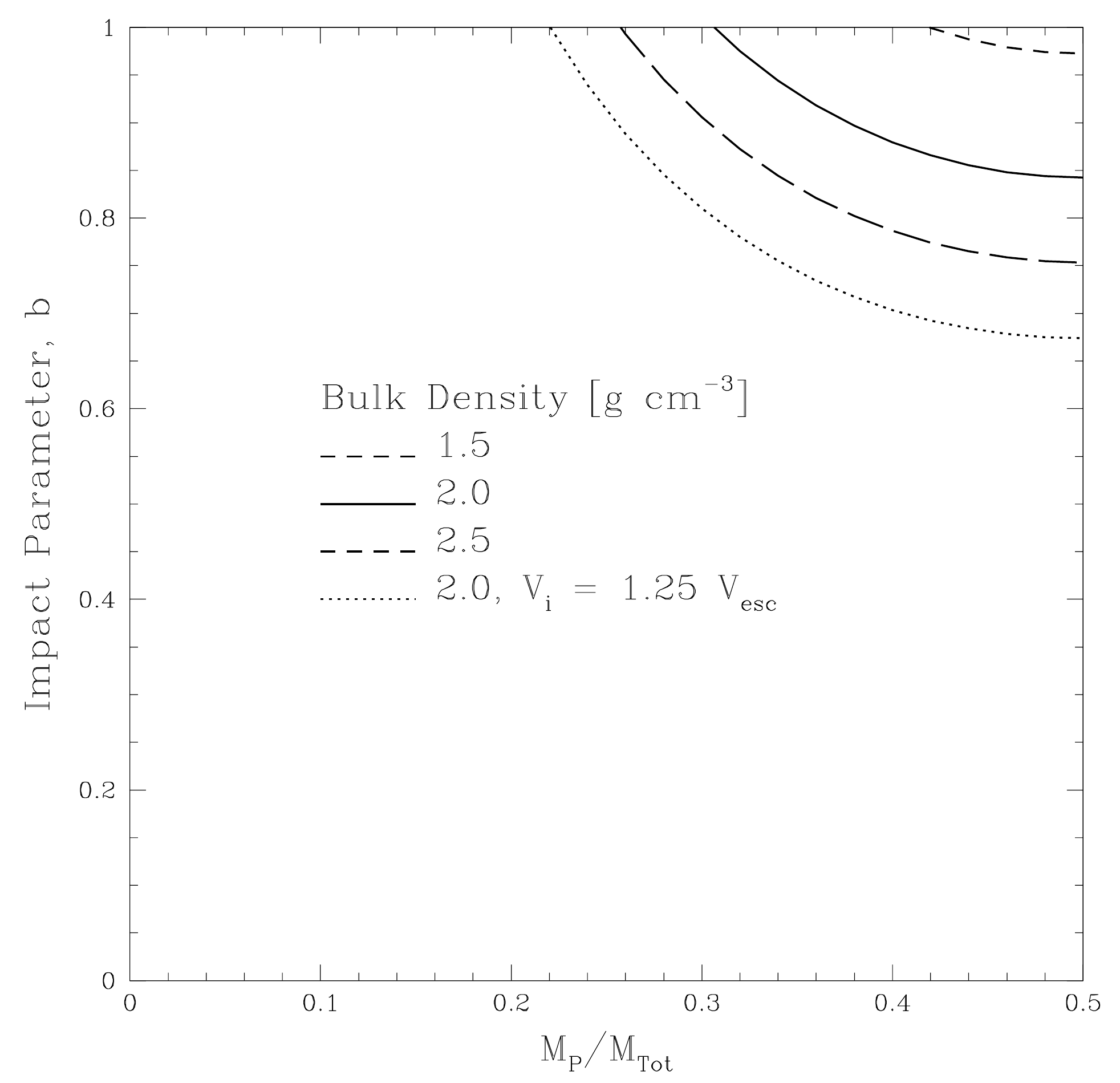}
 \caption{Analytic prediction of impact scenarios leading to the
   angular momentum of Haumea for various bulk densities and $V_i$. Solutions are
   lines as a function of impact parameter, $b$, and mass of the
   projectile normalized by the total mass,
   ${\rm M_P}/\mathrm{M_{Tot}}$. For $V_i = V_{\rm esc}$
   (Eq.~\ref{eq:final}) the short dashed, solid, and long dashed lines show solutions for bulk densities of 1.5, 2.0, and 2.5 g cm$^{-3}$, respectively. For $V_i = 1.25 V_{\rm esc}$ (Eq.~\ref{eq:bvi}) the dotted line shows solutions for a bulk density of 2.0 g cm$^{-3}$.  Haumea parameters
  are taken from
   \cite{Rabinowitz2006}.
   \label{fig:analytic}}
\end{figure}

For Haumea we assume a mass of $4.2\times 10^{21}$ kg and a spin period 3.92 hr from \citet{Rabinowitz2006}. Using these values, $\frac{L}{L_{\mathrm crit}}$ ranges between 1.1 and 0.8 for a plausible range of bulk densities of the colliding bodies (1.5 to 2.5 g cm$^{-3}$). The results from Eq.~\ref{eq:final} and \ref{eq:bvi} are shown in Fig.~\ref{fig:analytic}.

If the impact velocity equals the mutual escape velocity of the
projectile and target, Eq.~\ref{eq:final}
requires an impact parameter close to one ($> 0.8$) and a projectile
close to the mass of the target in order to attain the angular
momentum in Haumea (Fig.~\ref{fig:analytic}). Increasing the bulk
density of the bodies broadens the range of impact parameters that
could transfer the observed angular momentum. Similarly, when $V_i >
V_{{\rm esc}}$, there is a larger range of possible
projectile-to-target mass ratios that could produce the angular
momentum of Haumea (Eq.~\ref{eq:bvi}, dotted line in
Fig.~\ref{fig:analytic} assumes the mean bulk density for large KBOs
of 2~g~cm$^{-3}$). This work assumes no initial spin in the target and projectile. Initial spin in the same sense as the spin angular momentum would potentially increase the range of mass ratio that could produce a Haumea-like remnant.

As the impact velocity increases, the remaining body does not retain
most of the angular momentum of the encounter, as is assumed in the
equations above. When the impact velocity is large enough to begin to
disrupt the target \citep[3 -- 3.5 km s$^{-1}$,][]{Stewart2009}, a significant amount of angular momentum will be
carried away by the smallest fragments \citep[see Fig.~2 in][]{Leinhardt2000}.
Compensating for the
partial loss of angular momentum by further increasing the impact
velocity will lead to the catastrophic disruption regime. Recall that
the catastrophic disruption regime is ruled out based on the velocity
dispersion among the family members. Although the angular momentum
distribution in the catastrophic regime has not been extensively
studied, we expect that angular momentum transfer to the largest
remnant is inefficient. The largest remnant is a gravitationally
reaccumulated body; hence, the angular
momentum of the reaccumulated mass does not approach the spin
instability limit. Numerical simulations of catastrophic
disruption that resolve the shape of the largest remnant produce
spherical remnants not fast spinning elongated remnants \citep{Leinhardt2000,Leinhardt2009}.

Thus, the analytic argument suggests that the most straightforward way to create a body with high angular
momentum via a collision is in a grazing impact between two objects of
similar size at a velocity close to the mutual escape speed. Note, however, that the analytic solution makes
several assumptions. For example, mass that is lost from the system as
a result of the collision is not taken into account. The escaping mass
carries away some energy and momentum. In addition, the energy and
momentum conservation are approximated and do not include terms for
energy lost in heat and fracturing of the target and projectile.

 \subsection{Low Resolution Simulations} \label{sec:lowres}
To examine the predictions of the idealized analytic solutions, we
conducted a series of low-resolution simulations of impacts between
gravitational aggregates (Table~\ref{tab:pkd}). Because the encounter
velocities are modest ($\le 1.2$~km~s$^{-1}$), the energy lost to
shock deformation during the collision is minimal, and gravitational
forces dominate. Thus, for efficiency, the low-resolution simulations
utilized the $N$-body gravity code \texttt{pkdgrav}, which resolves
inelastic particle--particle collisions
\citep{Richardson2000,Leinhardt2000,Leinhardt2002}. Particle
collisions were modeled using a hard sphere model, where the
unbreakable spherical particles are non-penetrating.  The outcome of
each inelastic collision were governed by conservation of momentum and
typical coefficients of restitution of 0.5 and 1.0 in the normal and
tangential directions, respectively, for particles representing ice or
rock \citep[see discussion in][]{Leinhardt2009}.

Each projectile and target was modeled as a rubble pile, a
gravitationally bound aggregate of 955 particles with no tensile
strength \citep{Leinhardt2000,Leinhardt2002}.  Previous simulations
\citep{Leinhardt2000} show that a thousand particles is enough to
resolve general shape features in a rubble pile.  We assumed two
internal configurations: homogeneous and differentiated bodies. In the
differentiated cases, the colliding bodies had two layers: a 1.0 g
cm$^{-3}$ mantle representing ice and a 3.0 g cm$^{-3}$ core
representing rock. The mass ratio of the icy mantle to rocky core was
also varied to reach the desired bulk density, which resulted in a
range of initial radii from about 500 to 800 km.

Based on the analytic solutions, the projectile-to-target mass ratio
was assumed to be one in most cases with a subset of simulations
considering $M_{\rm P} / M_{\rm T} = 0.5$. The parameter space
included initial bulk densities from 1.0 to 3.0 g cm$^{-3}$, impact
velocities between 0.7 to 1.2 km s$^{-1}$, impact parameters from 0.55
to 0.71, and total system masses of $4.5$ to $8.2\times 10^{21}$ kg.

Due to the low resolution, each individual particle had relatively large mass, and it was difficult to strip material from the surface of the largest remnant. As
a result, we use these simulations only to refine the impact
parameters that reproduce the observed rotation period and approximate mass of Haumea. 
The calculated mass of the
largest remnant will be biased slightly upward, and the family members
will be completely unresolved.

\subsection{High Resolution Simulations \label{sec:gadget}}

Based on the low-resolution simulation results, impact parameters were
chosen for high-resolution calculations of the formation of Haumea and
its family members (Table \ref{tab:sims}).  We used a hybrid hydrocode
to $N$-body code technique similar to the method used in studies of
catastrophic collisions in the asteroid belt and the outer solar
system \citep[eg.][]{Michel2003,Durda2004,Nesvorny2006,Leinhardt2009}. The hybrid technique captures the
shock deformation during the early stage of the collision and follows
the gravity-controlled evolution of the material to very late
times. We used GADGET \citep{Springel2005}, a smoothed particle
hydrodynamics code (SPH) modified to use tabular equations of state
\citep{Marcus2009}, for the hydrocode phase and \texttt{pkdgrav} for the gravity
phase of the calculation. 

SPH is a Lagrangian technique for solving the hydrodynamic equations in which
the mass distribution is represented by spherically symmetric
overlapping particles that are evolved with time \citep{Gingold1977,
  Lucy1977}.  SPH has been used extensively to model impacts in the solar system from
asteroid collisions and family forming events to the formation of the
Pluto-Charon system \citep[e.g.][]{Asphaug1998,Michel2003,Canup2005}. Although GADGET includes self-gravity it is not practical to use a hydrocode for the entire integration of the collision as the timestep is limited by the Courant condition. 

The targets and impactors were differentiated bodies composed of an
ice mantle over a rock core with a bulk density of $\sim
2$~g~cm$^{-3}$. The largest KBOs have bulk
densities of about 2~g~cm$^{-3}$, which is similar to the density
predicted from cosmochemical estimates of the rock to ice ratio in the
outer solar system \citep{McKinnon1997,McKinnon2008}. Although the
internal structures of large KBOs are unknown \citep{Leinhardt2008},
strong water features on the surfaces of the largest bodies suggest
that they have differentiated \citep{McKinnon2008}. Hence, in this
work, we consider collisions between differentiated bodies only in the
high resolution simulations. 

The material in the rocky cores were modeled using a tabulated version
of the molecular ANEOS equation of state for SiO$_{2}$
\citep{Melosh2007}, and the ice mantles were modeled with the tabular
5-Phase equation of state for H$_{2}$O \citep{Senft2008}.  The
internal temperature profile is dependent on the ice to rock ratio and
the viscosity of ice; models indicate that temperatures for a Charon
size body are generally low \citep[e.g., $<300$~K after
4~Ga][]{McKinnon2008}. As a result, a constant initial temperature of
150~K was chosen. The bodies were initialized with hydrostatic
pressure profiles. They were then allowed to settle in isolation for many dynamical times at which point all particles have negligible velocities (cms s$^{-1}$) at the specified temperature. The
number of particles ranged from $\sim 1.2 \times 10^{5}$ to $\sim 4
\times 10^{5}$ ($\sim$ 24-36 particles per target radius) a sufficient resolution to
resolve shock heating and the formation of family members, and the
results were checked for sensitivity to resolution. As in all previous
studies of giant impacts, the materials are hydrodynamic (see
discussion in \S \ref{sec:cth}).

The GADGET simulations were run as long as was practically feasible,
normally about 60 simulation hours. At this time, the collision and
mass loss process was complete; however, more time was needed to
determine the long-term orbital stability of material bound to the
largest object.

\subsubsection{Orbital Evolution of Collision Fragments}

The N-body code \texttt{pkdgrav} was used to integrate the orbiting
fragments for thousands of spin periods of the largest remnant. 
The GADGET output was translated and handed off to \texttt{pkdgrav}. In
previous work on asteroid family formation, pairs of \texttt{pkdgrav}
particles were merged into a single particle after each particle-particle collision to
reduce computation time (resulting in artificial perfectly spherical collision remnants). In this work, because of the significant
elongation in the largest remnant, the shape and rotation rate needed
to be preserved for the orbital evolution calculation, and particle
merging could not be used. Hence, the \texttt{pkdgrav} calculation
utilized inelastic collisions (\S \ref{sec:lowres}) to preserve the shape
and gravitational potential of the largest remnant. 

However, the number of particles in the largest remnant in the GADGET
simulation ($\sim 10^5$ particles) is too large to integrate in
\texttt{pkdgrav} because of the computational expense of calculating
the collisions within the largest remnant.  Therefore, the largest
remnant was de-resolved to contain $\sim 10^3$ particles by placing a
grid over the body and placing all particles within a grid cell into a
single particle. Each merged particle had a mass equal to the combined
mass, a position equal to the center of mass position, and velocity
vector equal to the center of mass velocity.  The spin and shape of
the largest remnant was preserved.  The mass of the largest remnant
using inelastic collisions was compared with a perfect merging
collision outcome to test the stability of the handoff. The mass of
the largest remnant was very similar in both cases.

The shape and ice-to-rock ratio of individual smaller remnants (the
satellites and family members) were not resolved in the SPH
simulation. Hence, the self-gravitating remnants outside of the
largest remnant were merged into single particles. In this mannner,
the de-resolved \texttt{pkdgrav} calculation allowed the stability
simulation to run for thousands of orbits in a reasonable amount of
time.

\section{Results} \label{sec:results}

\subsection{A Collision Scenario for the Haumea System}\label{sec:pkdres}

\begin{deluxetable*}{ccccccccccccc}
\tabletypesize{\footnotesize}
\tablecaption{Summary of parameters and results from selected pkdgrav simulations.\label{tab:pkd}}
\footnotetext{$R$ -- radius of target; $\frac{M_{\rm P}}{M_{\rm T}}$
  -- mass of projectile normalized by mass of target; $\rho_b$,
  $\rho_c / \rho_m $ -- bulk density, density ratio of core to mantle;
  for $\rho_c / \rho_m > 1.0$, $\rho_c = 3.0$ and $\rho_m = 1.0$
  g~cm$^{-3}$; $b$ -- impact parameter; $V$ -- first impact velocity;
  $M_{\rm lr} / M_{\rm Tot}$ -- mass of largest remnant normalized by
  total mass; $\rho_{\rm lr}$ -- bulk density of lr; $P_{\rm lr}$ --
  spin period of lr. In all cases, each body contained 955 particles
  and $M_{\rm Tot}=4.5 \times 10^{21}$~kg.  Bulk density, calculated
  by circumscribing all particles within an axisymmetric ellipsoid, is
  always a minimum value. }
\tablehead{
  $R$& $\underline{M_{\rm P}}$&$\rho_b$ & $\underline{\rho_c}$ & $b$ &
  $V$ & $\underline{M_{\rm lr}}$ & semi-axes of lr & $\rho_{\rm lr}$ & $P_{\rm lr}$ & Collision Type\\
  km & $M_{\rm T}$ & g/cm$^{3}$ & $\rho_m$ & -- & km/s &$M_{\rm Tot}$ & km$\times$km$\times$km & g/cm$^{3}$ & hr & }
\startdata
  770 & 0.5 & 2.0 & 1.0 & 0.80 & 0.7 & 0.98 & 1270$\times$750$\times$690\footnote{lr is significantly non-axisymmetric.}& 1.6 & 4.4 & merge \\
  770 & 0.5 & 2.0 & 1.0 & 0.80 & 0.8 & 0.88 & 1000$\times$708$\times$670& 2.0 & 5.0 & graze \& merge \\
  770 & 0.5 & 2.0 & 1.0 & 0.80 & 0.9 & & & & & graze \& run \\
  650 & 1.0 & 2.0 & 1.0 & 0.65 & 0.8 & 0.99 & 1396$\times$711$\times$632 & 1.7 & 4.9 & merge\\
  650 & 1.0 & 2.0 & 1.0 & 0.65 & 0.9 & 0.99 & 1398$\times$750$\times$670 & 1.5 & 5.0 & graze \& merge\\
  650 & 1.0 & 2.0 & 1.0 & 0.65 & 1.0 &  & & & & graze \& run\\
  800 & 1.0 & 1.0 & 1.0 & 0.6 & 1.0 & & & & & graze \& run \\
  650 & 1.0 & 2.0 & 3.0 & 0.6 & 1.0 & & & & & graze \& run \\
  630 & 1.0 & 2.6 & 3.0 & 0.6 & 1.0 & 0.99 & 1270$\times$657$\times$607 & 2.1 & 3.8 & graze \& merge \\
  580 & 1.0 & 3.3  & 3.0 & 0.6 & 1.0 & 0.99 & 1164$\times$606$\times$526 & 2.8 & 3.7 & graze \& merge \\
  650 & 1.0 & 2.0 & 3.0 & 0.55 & 0.7 & 0.99 & 1219$\times$717$\times$699 & 1.8 & 4.1 & merge\\
  650 & 1.0 & 2.0 & 3.0 & 0.55 & 0.8 & 0.99 & 1293$\times$692$\times$666 & 1.8 & 4.3 & merge\\
  650 & 1.0 & 2.0 & 3.0 & 0.55 & 0.9 & 0.99 & 1385$\times$680$\times$634 & 1.8 & 4.2 & merge\\
  650 & 1.0 & 2.0 & 3.0 & 0.55 & 1.0 & 0.99 & 1362$\times$708$\times$647 & 1.7 & 4.5 & graze \& merge\\
  650 & 1.0 & 2.0 & 3.0 & 0.55 & 1.1 & & & & & graze \& run \\
  650 & 1.0 & 2.0 & 3.0 & 0.55 & 1.2 & & & & & graze \& run \\
  650 & 1.0 & 2.0 & 3.0 & 0.60 & 0.8 & 0.99 & 1303$\times$702$\times$688 & 1.7 & 4.1 & merge\\
  650 & 1.0 & 2.0 & 3.0 & 0.60 & 0.9 & 0.99 & 1322$\times$446$\times$636 & 1.7 & 3.9 & graze \& merge \\
  650 & 1.0 & 2.0 & 3.0 & 0.60 & 1.0 & & & & & graze \& run \\
  650 & 1.0 & 2.0 & 3.0 & 0.65 & 0.7 & 0.99 & 1252$\times$740$\times$672 & 1.7 & 4.1 &  merge \\
  650 & 1.0 & 2.0 & 3.0 & 0.65 & 0.8 & 0.99 & 1273$\times$698$\times$697 & 1.7 & 4.2 & graze \& merge \\
  650 & 1.0 & 2.0 & 3.0 & 0.65 & 0.9 & 0.99 & 1445$\times$699$\times$648 & 1.6 & 4.9 & graze \& graze \& merge\\
\enddata
\end{deluxetable*}

The impact conditions and outcomes of the low-resolution numerical
simulations of possible Haumea-forming impact events are summarized in
Table~\ref{tab:pkd}. The last column of the table, collision type,
describes the general class of the collision. We found three different
collision outcomes in our restricted parameter space (see \S \ref{sec:disc} for further discussion): 1) {\it merge}
-- the projectile and target merge after initial impact with little or
no mass loss; 2) {\it graze \& merge} -- the projectile initially hits
the target with a large impact parameter and then separates, the
projectile is decelerated, but remains relatively intact, and subsequently recollides at a much slower
velocity resulting in a merger and a fast-spinning body; 3) {\it graze
  \& run} -- the projectile and target hit but do not lose enough
energy to remain bound to each other. 

The impact parameters that produced elongated bodies with a total mass
and spin period similar to Haumea tend to be of the graze and merge
category. Thus, collisions that form a Haumea-like body are found in a
distinctly different parameter space than catastrophic disruption
events.

\begin{figure*}
\includegraphics[scale=0.5]{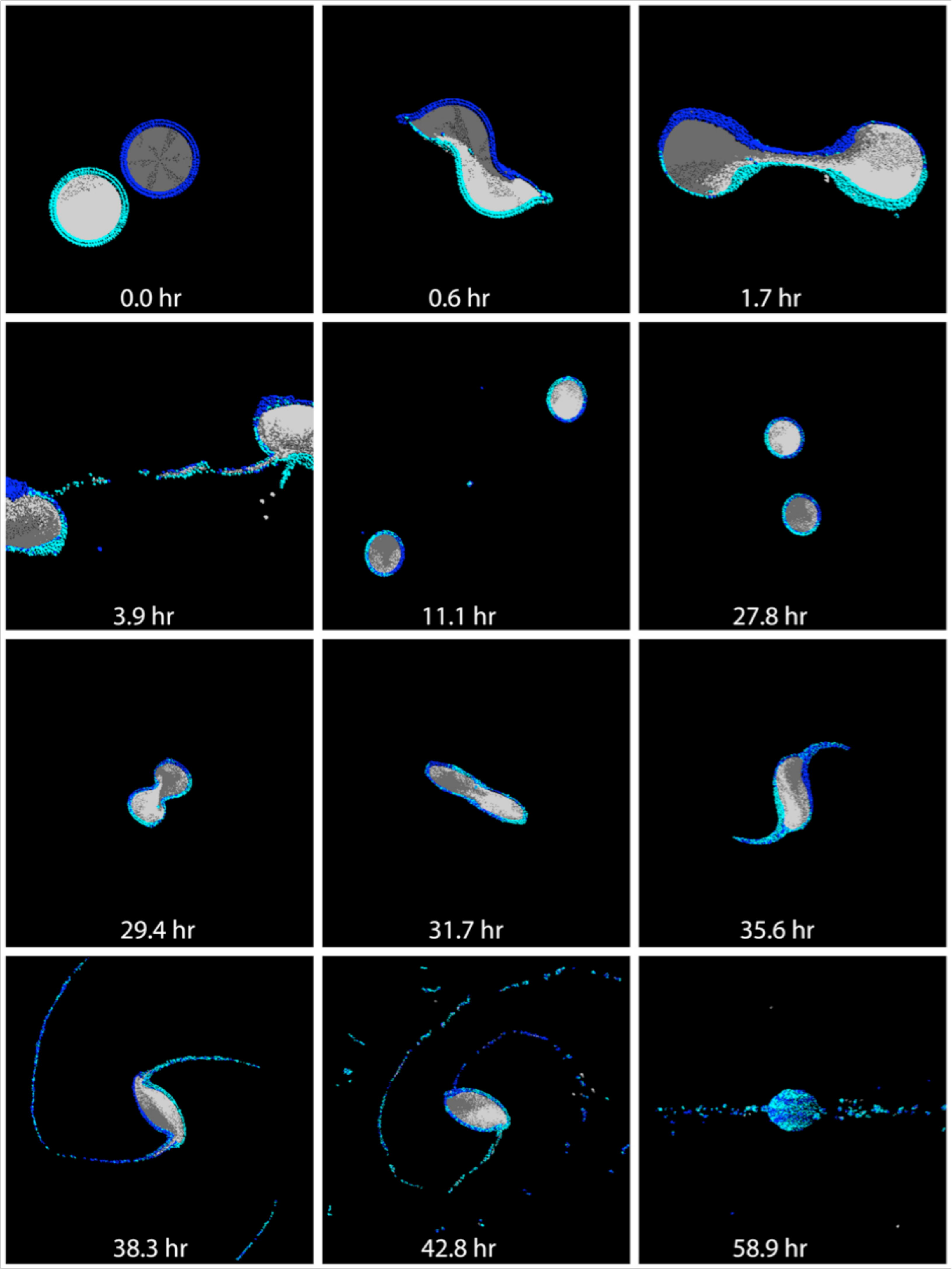}
\caption{Time series of a graze and merge event: 650-km diameter
  bodies colliding at 900 m~s$^{-1}$ with an impact parameter of 0.6
  (sim.~4 in Table \ref{tab:sims}). Cross section view through the collision plane which is in the page. Field of view is
  initially 5000$\times$5000~km, increasing to 10000$\times$10000~km
  at 11.1~hours.  The last frame (58.9 hrs) shows the system edge
  on. Color denotes the provenance of the materials: icy mantles (cyan
  and blue) and rocky cores (light and dark grey). For visual clarity in the last frame, 
  the radius of each dot equals one fifth of the smoothing length of
  the corresponding SPH particle, for all other frames the dots are point particles. Although some material is exchanged during the
   first impact, each body remains largely intact after
   separation. The rocky cores merge after the second impact, forming
   a differentiated primary. The surface of the merged body has
  distinct patches of ice that originate from each of the precursor
  bodies. The fragments thrown from the merged body are primarily
  material from the icy mantles. An animation is available in the
  online version of the Journal. \label{fig:tile}}
 \end{figure*}

High-resolution hybrid simulations of the most successful collision
scenarios for forming a Haumea-like planet were conducted to
investigate the properties of the satellites and family members
(Table~\ref{tab:sims}). A time series from an example collision simulation (sim.~4) is presented in Fig.~\ref{fig:tile}, which displays the materials in cross-section looking down on the collision plane. The last frame, which has been rotated by ninety degrees to show the collision outcome edge-on, shows the surfaces of the largest remnant and the debris field. During the collision, the rocky
cores of the progenitor bodies merge, and the resulting primary body
spins so quickly that it sheds icy mantle material from the ends in
many small clumps.  Some of this material is gravitationally bound and
some escapes from the primary. In this scenario, the satellites and
family members do not originate from the initial contact; instead,
they are spun off after the subsequent merger. As a result, the
V$_\infty$ of the family members are small; in other words, the
ejection velocities of the fragments are not much greater than the
escape velocity of the merged primary.

The analytic and numerical results show that the optimum parameter
space to form a Haumea-like planet is in an encounter slightly more
energetic than merging between comparably sized bodies.  Some of the
merging cases that are close to the transition to graze and merge also
produce an elongated, fast-spinning largest remnant; however, from
high-resolution simulations, we find that such merging simulations
eject less mass and, thus, do not tend to form families.  Although the long-term relaxation
of the bodies is not considered here, we note that impacts with a mass
ratio less than one produced initially nonsymmetric largest remnants
that are not consistent with observations.

The collision parameters that achieve the best agreement with
observations fall in a narrow parameter space: an impact speed of
800-900 m s$^{-1}$ and an impact parameter between 0.6 and 0.65 for
equal mass progenitors with bulk densities of 2~g~cm$^{-3}$.  At
higher impact velocities or higher impact parameters, the two bodies
escape from each other after the impact. At lower impact velocities or
lower impact parameters, the two bodies merge and less material is
ejected as potential family members.  Initial spin would increase or
decrease the impact parameter needed to achieve the same total angular
momentum, as shown in idealized cases in \cite{Leinhardt2000}.

 \begin{figure*} 
 {
 \includegraphics[scale=0.4]{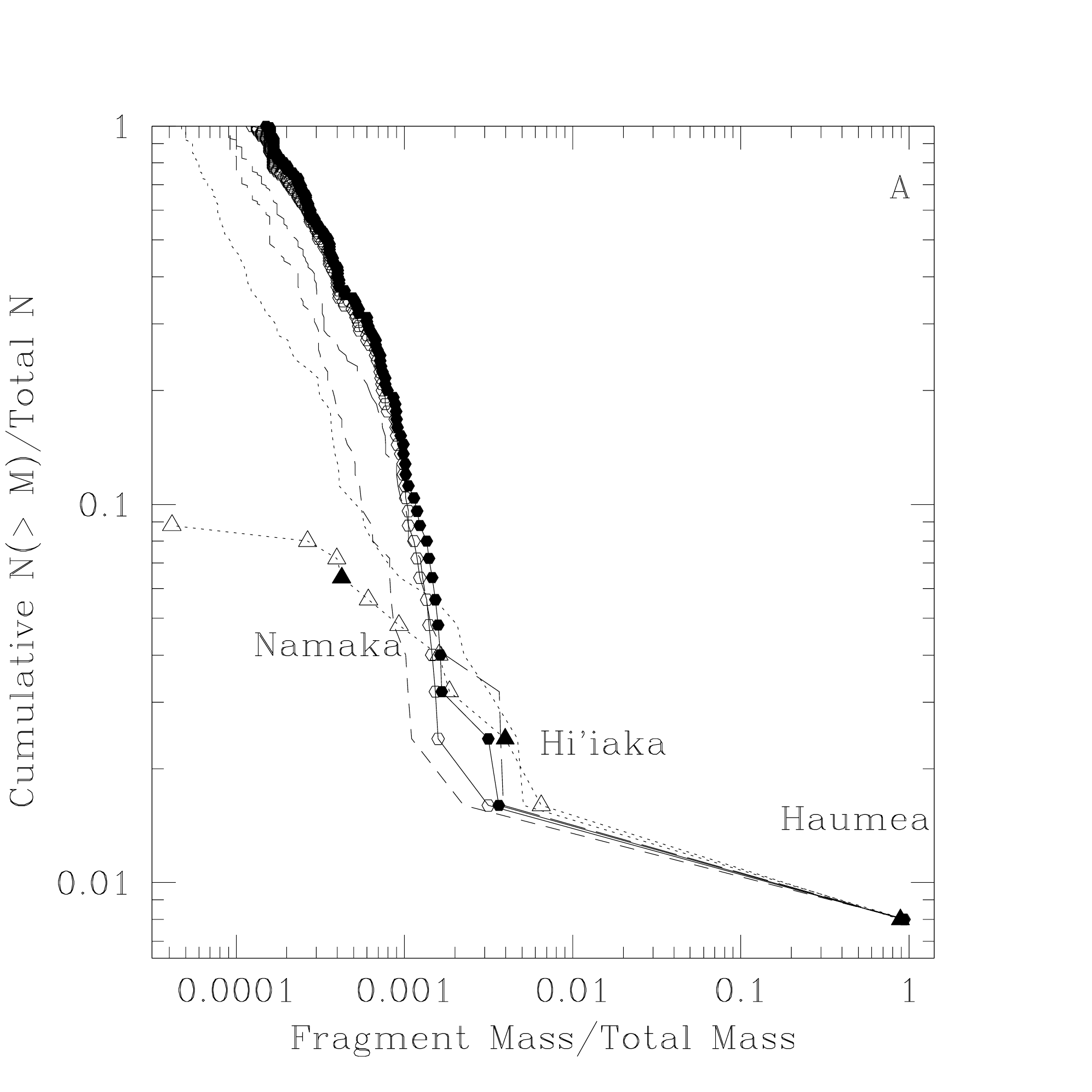}
     \includegraphics[scale=0.4]{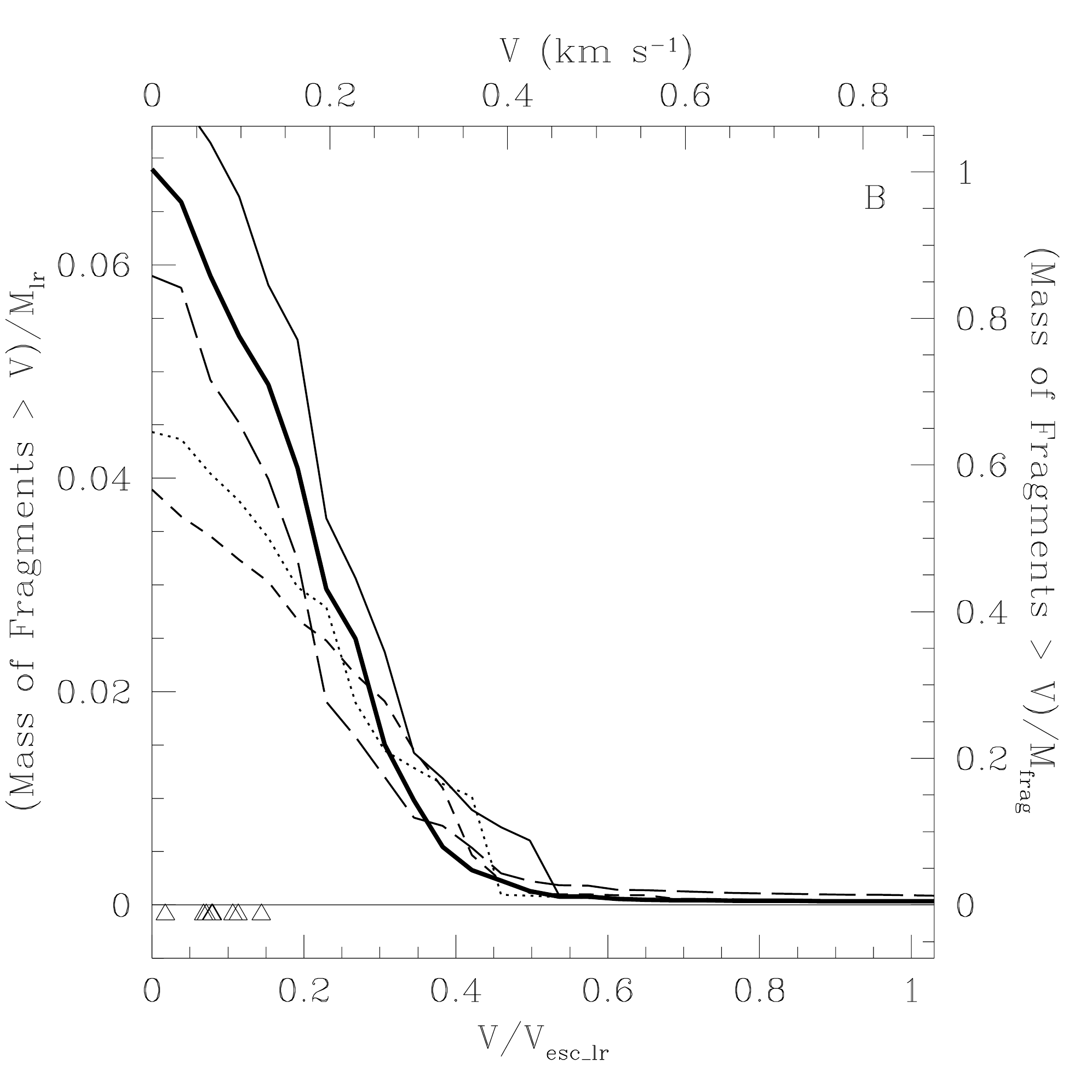}}
   \caption{Mass and velocity distributions for simulations 1-4 in
     Table \ref{tab:sims}. (A) Cumulative number of resolved fragments
     larger than mass M versus fragment mass (short dash, dotted, long dash,
     and filled circles are results from sim.~1-4 after $\sim$ 100
     orbits, the open circles are results from sim.~4  after $\sim$ 2000
     orbits). Triangles and dotted line are the observed family
     normalized to sim.~ 4. (B) Cumulative velocity distribution
     of fragments (short dashed, dotted, long dashed, and thin solid
     line, sim.~1-4, respectively, after 100 orbits, thick black
     line is from sim.~4 after 2000 orbits). The lower absicca is
     normalized by the escape speed from the largest remnant (model
     Haumea). The left ordinate is the mass of fragments with relative
     velocities greater than V normalized by the mass of the largest
     remnant, and the right ordinate is normalized by the total mass
     in smaller fragments in long running sim.~4. Open triangles
     below zero are minimum relative velocities of 
     family members. \label{fig:massdist}}
    \end{figure*}
 
\begin{deluxetable*}{cccccccccccccc}
  \tablecaption{Summary of Haumea family forming simulation results and
    observations.\label{tab:sims}}
  \footnotetext{$M_{\rm T}$ -- mass of target body; $R_{\rm T}$ --
    radius of target body; $\mu$ -- mass ratio of projectile to
    target; $b$ -- impact parameter; $V_{\rm imp1}$ -- first impact
    velocity; $V_{\rm imp2}$ -- second impact velocity in graze and
    merge event; $M_{\rm lr}$ -- mass of largest remnant; $\rho_{\rm
      lr}$ -- bulk density of lr; $P_{\rm lr}$ -- spin period of lr;
    $f_{\rm H_2O}$ -- ice mass fraction of all smaller
    fragments. $\rho_{lr}$ and $f_{\rm H_2O}$ from end of hydrocode
    phase (50--60 hrs); all other simulation results from end of
    gravity phase ($\sim 100$ orbits).}
\tablehead{
  Sim. & $M_{\rm T}$ & $R_{\rm T}$ &$\mu$ & $b$ & $V_{\mathrm{imp1}}$
  & $V_{\mathrm{imp2}}$ & $M_{\mathrm{lr}}$ &
  $\rho_{\mathrm{lr}}$ & semi-axes of lr & $P_{\rm lr}$ & $f_{\rm H_2O}$\\
  No. & $10^{21}$ kg & km & & & m s$^{-1}$ & m s$^{-1}$ & $10^{21}$ kg & g cm$^{-3}$ & km $\times$ km $\times$ km & hr & \\
}
\startdata    
  1 & 2.25 & 650 & 1 & 0.6 & 800 & 260 & 4.3 & 2.2 & $960 \times 870 \times 640$ & 3.6 & 0.86\\ 
  2\footnote{Higher resolution version of sim.~1 ($N=4\times10^5$).} & 2.25 & 650 & 1 & 0.6 & 800 & 240 & 4.3 & 2.1 & $1090 \times 820 \times 680$ & 3.4 & 0.79\\
  3 & 2.25 & 650 & 1 & 0.65 & 800 & 280 & 4.3 & 2.2 & $1100 \times 940 \times 640$ & 3.7 & 0.73 \\
  4 & 2.25 & 650 & 1 & 0.6 & 900 & 260 & 4.2 & 2.1 & $1100 \times 810 \times 620$ & 3.9 & 0.80\\
  5\footnote{Proposed impact scenario from \cite{Brown2007}.} & 4.80 & 830 & 0.22 & 0.71 & 3000 & - & 4.64 & 2.3 & $1700 \times 1500 \times 1500$ & 28 & - \\
  & \\
\enddata
  Obs.\footnote{Observed characteristics of Haumea
    \citep{Ragozzine2009,Rabinowitz2006}. Note: simulations presented
    here attempted to match the mass of $4.21 \pm 0.1 \times 10^{21}$
    kg quoted in  \citet{Rabinowitz2006}.} 
  & - & - & - & - & - & - & 4.006 & 2.6 & $1000 \times 750 \times 500$ & 3.9 & -\\

\end{deluxetable*}

\subsection{Properties of the Satellites and Family}

The impact conditions defined above reproduce the mass, spin period
and elongation of Haumea, as well as the mass and velocity
distributions of the observed satellites and family members.  In
Fig.~\ref{fig:massdist}A, the observed family \citep[triangles and dotted line from][]{Ragozzine2007} are normalized to the number of resolved
fragments in sim.~4 after 2000 spin orbits to facilitate
comparison. Unresolved fragments have too few particles ($N<10$) at handoff to be numerically resolved. Haumea and the two known satellites are indicated by
filled triangles. The diameter of Haumea family members was derived
from the absolute magnitude assuming similar albedo to Haumea of $\sim
0.7$, and the diameter was converted into mass assuming a bulk density
of 1~g~cm$^{-3}$ for a primarily ice composition. In Fig.~\ref{fig:massdist}B, the open triangles below zero indicate the
minimum relative velocities of the known family members with respect
to the center of the family using the reconstructed position for
Haumea \citep{Ragozzine2007}.

As shown in Fig.~\ref{fig:massdist}B, the total mass of potential
satellites and family members is small: $<0.07 M_{\mathrm{lr}}$ for
sim.~1-4 from Table~\ref{tab:sims}.  Although the modeled mass
is larger than the known satellites and family members ($\sim 0.01
M_{\mathrm{lr}}$), one does not expect that all of these bodies have
been observed or that all should survive to the present day. Hence,
the modeled mass of smaller bodies is consistent with the observed
Haumea system.  In addition, the masses of individual satellite and
family members are small, with most $<10^{-3} M_{\mathrm{lr}}$, which
is also in agreement with the observed family
(Fig.~\ref{fig:massdist}A).

Almost all of the family members have a relative velocity that is
$<0.5 V_{\mathrm{esc}}$ with respect to the primary, with 80-90\% of
the mass having a speed $<300$ m s$^{-1}$ and 35-60\% of the mass
having a speed less than 150 m s$^{-1}$. Note that 150~m~s$^{-1}$ is
the {\it minimum} velocity dispersion of the observed family; the true
values could be about two times larger when accounting for the unknown
orbital orientation \citep{Ragozzine2007}.  The model predicts that
negligible mass reaches the typical velocity dispersion among 100-km
size KBOs of about 1000~m~s$^{-1}$.  

As a resolution test one simulation was completed at higher resolution (no.~2, dotted line in Fig.~\ref{fig:massdist}, $N=4\times 10^5$ compared to $N=1.2 \times 10^5$ for simulations 1, 3 - 5) with the same collision parameters as sim.~1. The number and velocity dispersion of the remnants are consistent with the lower resolution simulations thus we are confident we have reached resolution convergence.

The modeled satellite and family members are
comprised almost entirely of icy mantle material (73-86\% by mass). The bulk
density of the largest remnant increases by $\sim 10\%$ over the
initial density of the progenitor bodies as a result of the
preferential stripping of the lower-density ice mantle
(Fig.~\ref{fig:tile}).  The modeled bulk density of the largest
remnant is within the range of uncertainty for the density of Haumea.
These results explain the shared water ice spectral feature of the
bodies in the Haumea system and the complete lack of bodies with a
similar water ice spectrum in the general (not dynamically associated)
KBO population around Haumea. Negligble mass was dispersed into the
background KBO population in the graze and merge family-forming
event. 

In one simulation (no.~4), the orbits of the satellites and family
members were integrated for 2000 spin orbits to assess the stability
of the newly formed system around the elongated primary. At this time,
there are $\sim 35$ objects gravitationally bound and in orbit about
the primary (the mass in orbit is about $0.01M_{\rm lr}$ and the mass
in family members is about $0.06M_{\rm lr}$). These bodies have
eccentricities below 0.9 and instantaneous orbits that do not
intersect the primary. The bound objects and instantaneous orbital
parameters were determined using the \texttt{companion} code \citep{Leinhardt2005}. 
Here, we do not assess the longer term
dynamical evolution of the collisional system. However, we demonstrate
that some bodies remain in stable orbits around the elongated,
fast-spinning primary well after the collision event. Unlike
previously thought \citep{Brown2006}, the formation of multiple small
satellites around Haumea does not require accretion in a massive disk.

\subsection{The Effect of Material Strength \label{sec:cth}} 

Simulations of giant impact events usually utilize a SPH code \citep[e.g., this work,][]{Benz1986,Benz1988, Canup2004,Canup2005,Benz2007, Marinova2008}. A Lagrangian SPH calculation has the advantages of an arbitrarily large spatial domain and efficiency in tracking
small fragments. In an Eulerian (grid-based) code, the spatial domain
must be decided upon in advance and tracking small fragments through
the mesh is computationally very expensive. In previous studies, giant
impact calculations have neglected material strength on the grounds
that self-gravitational forces and shock pressure gradients dominate
the problem. Most giant impact studies have focused on the end stages
of planet formation and considered hypervelocity impact events that
generate strong shock waves. Hence, it has been reasonable to neglect
material strength. For collisions between dwarf planets at subsonic
velocities, however, it is not obvious that strength can be neglected.

We conducted a few comparison three-dimensional simulations using the
Eulerian shock physics code CTH \citep{McGlaun1990} with adaptive mesh
refinement \citep{Crawford1999}. CTH has the option to include
self-gravitational forces using the parallel tree method of
\citep{Barnes1986}.  The simulations had a resolution of 31 to 42 km
(30 to 40 cells across each initial body). In the nominal simulations,
each CTH cell is comparable in physical size to a single GADGET
particle within the initial bodies, although the effective resolution
in CTH is slightly higher because of the differences in smoothing
lengths between the two codes. Tests at twice the resolution (in each
dimension) yielded similar results. The CTH calculations utilize the
same tabulated equation of state models as used in the GADGET SPH
simulations. Each body was initialized in gravitational equilibrium at
a constant temperature of 150~K.  The radii were 650~km, with an ice
mantle over a rock core such that the bulk density was
2~g~cm$^{-3}$. To allow for reasonable calculation times, material
in cells with a bulk density less than 0.01 g~cm$^{-3}$ was discarded
from the calculation. Hence, fragments that become the moons and
family members of Haumea are not modeled in CTH. Only the formation of
the primary (Haumea) is considered.

Some simulations were hydrodynamic (no shear strength) for direct
comparison to GADGET. Other simulations utilized a simple friction law
(the geological yield model in CTH) that represents friction in
fractured (damaged) ice \citep{Senft2008}: $Y_{0} + \mu P$, where
$Y_0=0.1$~MPa is the cohesion, $\mu=0.55$ is the friction coefficient,
and $P$ is pressure. The shear strength is limited to a maximum of
0.1~GPa. Shear strength is thermally degraded as the temperature
approaches 273~K. The tensile strength was 1.7~MPa. For simplicity,
the same strength model was used in both the ice and rock components.
The strength parameters are similar to models of sedimentary rocks
\citep[e.g.,][]{Collins2008}, which are signficantly weaker than
crystalline rocks. 

\begin{figure*}
\begin{center}
\includegraphics[scale=.9]{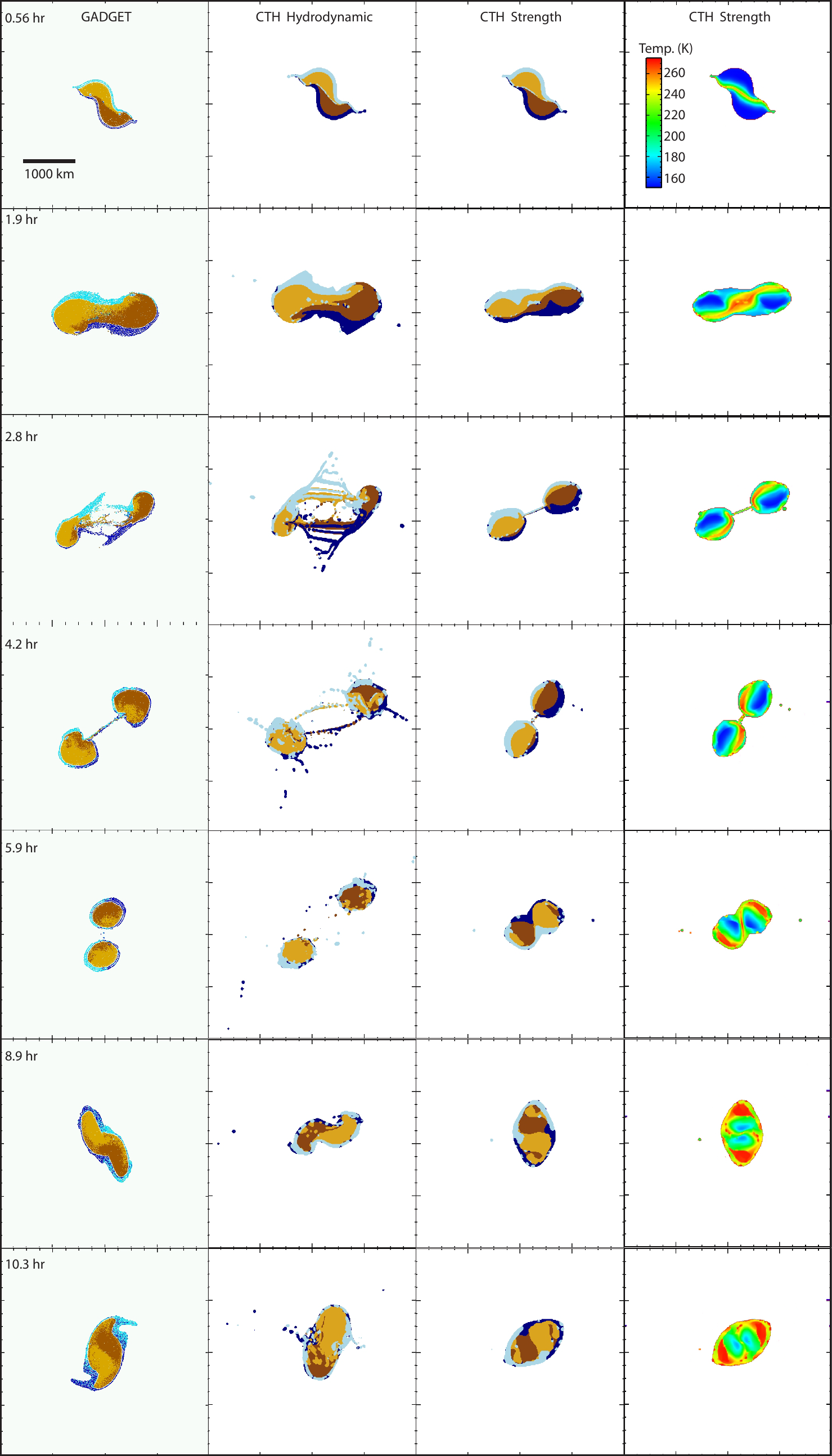}
 \end{center}
 \caption{
Time series of sim.~1 in Table~\ref{tab:sims} using GADGET
and CTH with and without strength. View of the cross section through the impact
plane. Each panel is $4000\times4000$~km. Columns 1-3: Colors
represent material (blues -- ice mantles; browns -- rock
cores). Column 4: Color represents temperature. 
\label{fig:cthcomp}}
\end{figure*}

Because the CTH calculations require significantly more computational
time than the GADGET calculations, comparisons between the two codes
were made at early stages in the impact event. Hydrodynamic CTH
calculations are similar to the GADGET results
(Fig.~\ref{fig:cthcomp}). The minor differences at late times are
due to small differences in the initial conditions: the initial
separation of the bodies and the better resolution of the ice-rock
interface in CTH.

In cases with strength, a number of differences are observed. The
radial oscillations in each body after the initial contact are not
observed with strength. These oscillations arise from momentum
transfer between the rock core and ice mantle; the difference in
particle velocities from the momentum transfer is damped with
strength.  The number of `strings of fragments' between the two bodies
after the initial contact is decreased with strength (2.8 hours in
Fig.~\ref{fig:cthcomp}). Because of dissipation of energy in
mechanical work during the first contact, the second contact occurs
sooner with strength. The temperature increase from the impact is
higher with strength because of dissipation of mechanical work as
heat. Note that with and without strength, the cores of the two bodies
merge quickly after the second contact. The purpose of the CTH
calculation is not to try to develop a realistic strength model for
large KBOs, but simply to demonstrate that the addition
of a reasonable amount of internal friction yields essentially the
same result as in the detailed hydrodynamic simulations.

\subsection{Impact-induced Heating and the Icy Surface \label{sec:temp}}

During the impact event, the temperature in the portions of the ice
mantle in the regions of contact are raised to the phase boundary
(melting or vaporization, depending on the local pressure, Fig.~\ref{fig:cthcomp}). However, negligible mass of ice is melted
or vaporized. For the $\sim$ 1~km~s$^{-1}$ impact velocities considered
here and a wide range of initial temperatures, ice that experiences
the peak shock pressures will be raised to the melting or vaporization
curve \citep{Stewart2008}. However, no ice is expected to completely
melt or vaporize. Thus, the outcome of the collision event is not
sensitive to the initial temperature as neither the rock nor ice
components experience signficant shock-induced phase changes during
the collision event. As a confirmation, a test SPH simulation with a
high temperature rock interior yielded essentially the same result.

In a graze and merge impact, the ice mantles experience severe
deformation that is expected to fragment the ice. Upon merging, solid
state differentiation occurs quickly, even with some strength. The
surface of the resulting primary (Haumea) is covered with ice
fragments that have been heated over a wide range: from negligibly to
the phase boundaries. The fragmented ice mantle has negligible
cohesion and may be likened to a slurry. Ice has exceedingly low friction when warm and at modest
slip velocities which are exceeded in the scenario considered here
\citep{Maeno2003}. Thus, we conclude that the ice mantle will have
negligible resistance to mass loss due to the high angular momentum of
the merged body.

Because the graze and merge portions of the impact event are similar
in the GADGET and CTH calculations (with and without strength), we
argue that the spin off of ice fragments at the end stages of the
event is reasonably modeled in the hydrodynamic GADGET calculations.

One of the remarkable characteristics of Haumea, its larger satellite,
and its family members is the strong water ice spectral feature. The
graze and merge family forming event explains the ice-dominated
surfaces of the Haumea system. 
Although the simulations cannot address
the details of ice separation from other phases, the results suggest
that the formation event produced an icy surface that is cleaner
compared to other KBOs, as is observed \citep{Pinilla2009}.  A
relatively clean icy mantle is necessary to prevent reddening from
cosmic irradiation \citep{Rabinowitz2008} over the time since the
impact event \citep[$>1$ Ga,][]{Ragozzine2007}. Because the surface of
Haumea appears so homogeneous and unlike other KBOs, we argue that it
is unlikely that the surfaces of the precursor bodies were so
similar. Hence, the strong ice feature on the Haumea family must be
related to the family-forming impact event.

\subsection{Additional Formation Scenarios}

\citet{Brown2007} suggested a possible impact scenario to
produce the Haumea collisional family based on numerical studies of
asteroidal family forming events \citep{Benz1999}: impact of a projectile with 0.22 the
mass of the target at 3 km~s$^{-1}$ at an impact parameter of
0.71. The authors suggested that such an impact would both strip off a
portion of the target's mantle and impart a high spin period. We conducted 
a high-resolution simulation of the proposed
impact scenario. Our results demonstrate that there is not enough
energy and momentum coupling between the projectile and the target to
produce a fast-spinning primary (sim.~5 in Table~\ref{tab:sims}).  The
projectile shaves off some of the target material and escapes from the
system, but the remaining angular momentum is insufficient to elongate
the target body. Based on our analytic calculations and numerical
results, it is not favorable for a small, fast projectile to impart
enough angular momentum to create a fast-spinning elongated body.

Using an order of magnitude analysis, \cite{Schlichting2009}
suggest a different Haumea family forming scenario involving a two
stage formation process. First, a giant impact creates an elongated
fast-spinning primary and a large, tightly-bound satellite. Second, a
subsequent impact onto the satellite, disrupts it and creates the
family members and small satellites. Based on numerical simulations,
an elongated and fast spinning primary is only produced in a slow
collision with a large impact parameter and a mass ratio close to
unity \citep[Table \ref{tab:pkd},][]{Leinhardt2000, Leinhardt2002}; furthermore, these scenarios do not
create a large tightly-bound satellite. Impact events that produce a
large tightly-bound satellite \citep[e.g.,][]{Canup2005} do not form
elongated primaries.

We have not completed an exhaustive parameter space study of family-forming collisions in the Kuiper Belt. It is possible that additional collisional scenarios such as graze and run could form the Haumea collisional family. Future work on this scenario would need to address the homogenization of the entire surface of Haumea (since it would not all melt as a result of the encounter) and the probabilities of the encounter providing enough angular momentum to a Haumea-sized target. In addition, a graze and run collision deposits only a small amount of internal energy into the target - not enough to differentiate the body. Here we present the first fully self-consistent formation model that does not require unusual pre-impact conditions.

\section{Discussion} \label{sec:disc}

Giant impacts are common in the late stages of planet formation, and
several outcomes are possible. With approximately decreasing impact energy, a giant
impact leads to: ({\it i}) {\it catastrophic disruption} \citep[half or more
of the target mass is lost,][]{Benz1999,Stewart2009,Marcus2009}. The largest remnants are
aggregates of gravitationally reaccumulated material; it is the
formation mechanism for asteroid belt families. ({\it ii}) {\it graze
  and run} \citep[a.k.a.~hit and run,][]{Asphaug2006}. In some oblique impacts, the projectile hits and then
escapes the target with both bodies remaining largely intact. Note that we prefer the term ``graze''
rather than ``hit'' to indicate the need for an oblique impact. 
({\it  iii}) {\it graze and capture}. The projectile obliquely hits the
target and separates, but does not have enough energy to escape and is
captured in orbit. It is the favored formation mechanism for the
Pluto-Charon binary \citep{Canup2005}.  ({\it iv}) {\it graze and
  merge}. The projectile obliquely hits, separates, and then
recollides and merges with the target. The high angular momentum of
the merged body spins off some material. Such a scenario explains the
unusual characteristics of the Haumea collisional family. In an
alternate scenario, a graze and merge impact to form Pluto may also
produce a disk of bound material of the mass of Charon
\citep{Canup2005}.  ({\it v}) {\it merge}. The projectile and target
merge during the first contact; in some cases, a small fraction of
material may be thrown into orbit or escape. The Earth's moon forming impact \citep[][]{Canup2004} falls in between the graze and merge and merging categories. The majority of the impact scenarios summarized above have left a distinct type of observable satellite or family system in the solar system.

The modeled impact scenarios that reproduce the Pluto and Haumea
systems are strikingly similar. The impact velocities, impact
parameters, and masses of the projectile are almost exactly the same,
with the primary difference being the mass ratio between the
projectile and target: 0.3 for Pluto \citep{Canup2005} and 1.0 for
Haumea. What is the liklihood of such impact events?  The Kuiper Belt
is presently composed of multiple dynamically distinct sub-populations
\citep[e.g., classical, scattered and resonant,][]{Morbidelli2008}.
There is strong evidence for an excitation event after an initial
period of collisional growth.  The Haumea family must have formed
after the excitation event, which would have scattered away all the
family members. Although collision probabilities between two 1000-km
scale bodies in the classical population are negligible,
\citet{Levison2008} calculated up to a $\sim 50\%$ probability of the
impact scenario proposed by \citet{Brown2007} if the two bodies
originated from the scattered population and then entered the
classical population as a result of the impact. However, the estimated
mean impact velocity between scattered objects is substantially higher
(about 2.7~km~s$^{-1}$) than we find for the Haumea-forming event. 
The probabilities of slower speed collisions need to be investigated.

\section{Conclusions}

In this work, we show that a graze and merge collision event between
nearly equal mass bodies is able to produce a symmetrically-elongated,
fast-spinning primary, a family of collisional fragments, and multiple
bound satellites. The satellites and family members are derived from
the icy mantle of the merged, differentiated primary. The family
members have a small relative velocity with respect to the
primary. This scenario matches all of the observed characteristics of
the Haumea system.  Our results predict that observations of future
family members should all have relative velocities less than $\sim 0.5
V_{\mathrm{esc}}$ of Haumea and that the family members should not be
isotropically distributed because they formed in a plane.

We now have the tools to read the record of giant collisions throughout the solar system. At present, there are several 10's of Pluto-sized bodies known in the
outer solar system \citep{Brown2008}. The new Pan-STARRS observatory will detect up to an order of magnitude more bodies in the Kuiper Belt \citep{Trujillo2008}, which will complete the Haumea system and test the predictions of the graze and merge scenario. Models of the dynamical
evolution of the Kuiper Belt indicate that the population must have
been much larger in the past (by a factor of e.g. $\sim 1000$)
\citep{Morbidelli2008}.  Satellite formation was a common outcome of
giant impacts, and the abundance of satellites around dwarf planets
indicates that giant impacts were frequent. We find that the types of
collision events that formed the observed satellites and families in
the outer solar system are distinctly different from the Earth's moon
formation and families in the asteroid belt. The narrow range of
impact parameters that formed the Pluto and Haumea systems place
strong constraints on the dynamical history of the largest bodies in
the Kuiper Belt.

\acknowledgements {\it Acknowledgements.} We thank D. Ragozzine and M. \'{C}uk for helpful discussions. The hydrocode calculations were run on the Odyssey cluster supported by the Harvard FAS Research Computing Group. The N-body calculations were run using the University of Cambridge, Astrophysical Fluids Research Group computational facilities. ZML is supported by a STFC fellowship; STS by NASA grant \# NNX09AP27G.

\newpage
\bibliography{scibib}

\end{document}